\documentclass[prl,twocolumn,showpacs,preprintnumbers,amsmath,amssymb,floatfix]{revtex4}

\usepackage{subfigure,graphicx,epsfig,amsmath,amsfonts,amssymb,xcolor,slashed,ulem,multirow}

\newcommand{\be}{\begin{equation}}
\newcommand{\ee}{\end{equation}}
\newcommand{\ba}{\begin{eqnarray}}
\newcommand{\ea}{\end{eqnarray}}

\newcommand{\mev}{\textrm{ MeV}}

\newcommand{\br}{{\rm BR}}

\newcommand{\minv}{M_{\rm inv}}
\newcommand{\minvpa}{M_{\rm inv}(\pi^0a_0)}
\newcommand{\minvpf}{M_{\rm inv}(\pi^0f_0)}
\newcommand{\sumpol}{\sum_{\rm pol}}
\newcommand{\dspkks}{D_s^+\rightarrow\pi^+K^+K^{*-}}
\newcommand{\dsppa}{D_s^+\rightarrow\pi^+\pi^0a_0(980)}
\newcommand{\ksb}{\bar{K}^*}
\newcommand{\ksbz}{\bar{K}^{*0}}
\newcommand{\ksm}{K^{*-}}
\newcommand{\kb}{\bar{K}}
\newcommand{\omkz}{\omega_{K^0}}
\newcommand{\omkszb}{\omega_{\bar{K}^{*0}}}
\newcommand{\omkzb}{\omega_{\bar{K}^0}}
\newcommand{\polkszb}{\vec{\epsilon}_{\bar{K}^{*0}}}
\newcommand{\polksm}{\vec{\epsilon}_{K^{*-}}}
\renewcommand\sout{\bgroup \color[rgb]{1,0,0} \ULdepth=-.5ex \ULset}

\begin{document}

\title{Abnormal isospin violation and $a_0 -f_0$ mixing in the $D_s^+ \to \pi^+ \pi^0 a_0(980) (f_0(980))$ reactions}
 
\author{S. Sakai}
%\email{}

\affiliation{Departamento de
F\'{\i}sica Te\'orica and IFIC, Centro Mixto Universidad de
Valencia-CSIC Institutos de Investigaci\'on de Paterna, Aptdo.
22085, 46071 Valencia, Spain}

\author{W. H. Liang}

\affiliation{Department of Physics, Guangxi Normal University, Guilin 541004, China}

\author{E. Oset}

\affiliation{Departamento de
F\'{\i}sica Te\'orica and IFIC, Centro Mixto Universidad de
Valencia-CSIC Institutos de Investigaci\'on de Paterna, Aptdo.
22085, 46071 Valencia, Spain}

\date{\today}

\begin{abstract} 
We have chosen the reactions $D_s^+ \to \pi^+ \pi^0 a_0(980) (f_0(980))$  investigating the isospin violating channel $D_s^+ \to \pi^+ \pi^0 f_0(980)$. The reaction was chosen because by varying the $\pi^0 a_0(980) (f_0(980))$  invariant mass one goes through the peak of a triangle singularity emerging from  $D_s^+ \to \pi^+\bar K^* K$, followed by  $\bar K^* \to \bar K \pi^0$ and the further merging of $K \bar K$  to produce the  $a_0(980)$ or  $f_0(980)$. We found that the amount of isospin violation had its peak precisely at the value of the 
 $\pi^0 a_0(980) (f_0(980))$  invariant mass where the singularity has its maximum, stressing the role of the triangle singularities as a factor to enhance the mixing of the 
 $f_0(980)$ and  $a_0(980)$ resonances. We calculate absolute rates for the reactions and 
show that they are within present measurable range. The measurement of these reactions would bring further information into the role of triangle singularities in isospin violation and the $a_0 -f_0$ mixing in particular and shed further light into the nature of the low energy scalar mesons. 
\end{abstract}

\maketitle

\section{Introduction}
The issue of the $f_0(980)-a_0(980)$ mixing has attracted much attention in the hadron community due to its potential to learn about the nature of the low lying scalar mesons.
First suggested in  Ref.~\cite{Achasov:1979xc}, it was very early identified as being tied to the mass difference between the charged and neutral kaons \cite{Achasov:1979xc,crisreport}.
Different reactions were suggested to find signals of this mixing in the
$pn \to d \eta \pi^0$ \cite{Kudryavtsev:2001ee}, the $\gamma p \to p
\pi^0 \eta$ \cite{Kerbikov:2000pu} and the $\pi^- p \to \pi^0 \eta n$
\cite{Achasov:2003se}. Finally, it was the $J/\psi \to \phi \eta \pi^0$
reaction
%the one that
which showed clearly a mixing. This reaction had been suggested in Ref.~\cite{Wu:2007jh} and estimates were done there. A more detailed calculation was presented in Ref.~\cite{Hanhart:2007bd}, using the chiral unitary approach \cite{npa,ramonet} to account for the interaction of pseudoscalar mesons that generate the $f_0(980)$ and $a_0(980)$ resonances, and the mechanism for $f_0(980)$ production used in Ref.~\cite{Meissner:2000bc} for the $J/\psi \to \phi \pi \pi$ reaction. In that paper the role of the $K \bar K $ loops and of the difference of masses between the $K^+$ and $K^0$ was further investigated. A further revision of this issue was done in Ref.~\cite{Roca:2012cv}, where the production model for  $J/\psi \to \phi \pi \pi$ and $J/\psi \to \phi \pi \eta$ was improved taking the more complete model of Ref.~\cite{Roca:2004uc} for $J/\psi \to \phi \pi \pi$. The work of Ref.~\cite{Roca:2012cv} reproduced very accurately the shape and magnitude of the $a_0(980)$ production in the $J/\psi \to \phi \pi \eta$ reaction \cite{Ablikim:2010aa}, together with the $f_0(980)$ production in $J/\psi \to \phi \pi \pi$ with no more free parameters than the one used to regularize the loops in the study of the pseudoscalar-pseudoscalar interaction in Ref.~\cite{npa}. The new mechanisms used in Ref.~\cite{Roca:2012cv}, accounting for sequential vector and axial-vector meson exchange, were found to be crucial in order to obtain the actual shape and strength (in about a factor of two) of the mass distributions. Further study of the mixing and suggestion of reactions
to observe it was done in Ref.~\cite{Wu:2008hx}.

   The concept of an $f_0(980)-a_0(980)$ mixing parameter was been accepted when a new reaction came to challenge it. The reaction was the $\eta(1405)$ decay to $\pi^0 f_0(980)$
measured at BESIII \cite{BESIII:2012aa}, which showed an unusually large isospin violation, or equivalently a very large $f_0(980)-a_0(980)$ mixing when compared with the isospin allowed $\eta(1405) \to \pi^0 a_0(980)$. This abnormal mixing found an explanation in Ref.~\cite{Wu:2011yx} due to the role of a triangle singularity involving a mechanism in which the $\eta(1405)$ decays to $K^* \bar K$ , followed by the decay of $K^*$ in $K \pi$ and the merging of $K \bar K $ to give the $f_0(980)$ or $a_0(980)$. Further work along these lines was done in Ref.~\cite{Aceti:2012dj} where ambiguities in the size of the $\eta(1405) \to \pi^0 a_0(980)$ in Ref.~\cite{Wu:2011yx} were solved. More work along these lines followed in Ref.~\cite{Wu:2012pg}, where it was suggested that the $\eta(1405)$ and  $\eta(1475)$ are actually the same state. 

   Triangle singularities (TS) were introduced by Landau \cite{landau} for the decay of an external particle and develop from a mechanism depicted by a Feynman diagram with three intermediate propagators. When the three intermediate particles are simultaneously placed on shell and are collinear in the rest frame of the decaying particle, a singularity can emerge if the process has a classical correspondence, which is known as the Coleman-Norton theorem \cite{Coleman:1965xm}. A modern and easy formulation of the problem is given in the paper \cite{guo}. 

   While finding physical examples was not successful at the origin of the formulation of the TS, the advent of vast experimental information nowadays is providing many examples of TS, sometimes simulating a resonance, other times providing mechanism for the production of particular modes in reactions.  Suggestions of places to look for triangle singularities were done in Ref.~\cite{qzhao}. One of them was the possibility that the COMPASS claimed "$a_1(1420)$" resonance \cite{Adolph:2015pws} would not be a genuine state but the manifestation of a TS with intermediate states $K^* \bar K K$. This hypothesis was made quantitative in Ref.~\cite{mikha}. The mechanism suggested implied the decay of the $a_1(1260)$ into $K^* \bar K$ , followed by the decay of $K^*$ into $K \pi$ and the further fusion of $K \bar K$ into the $f_0(980)$ giving rise to the decay mode $\pi f_0(980)$ observed in the experiment \cite{Adolph:2015pws}. Further refinements along this line with consideration of the $\rho \pi$ decay of the $a_1(1260)$ resonance, were done in Ref.~\cite{fcadai}, leading to a more accurate determination of the experimental observables and to the same conclusion. 

     Suggestions that the observed charged charmonium $Z_c(3900)$ \cite{Ablikim:2013mio,Ablikim:2013emm,Liu:2013dau,Xiao:2013iha} could be due to a TS were done in Ref.~\cite{qzhao,Wang:2013cya,Liu:2013vfa}, and similar claims were done regarding other quarkonium \cite{Liu:2013vfa} and bottomnium \cite{Liu:2014spa}. Claims that the narrow pentaquark state found by the LHCb collaboration \cite{Aaij:2015tga,chinese} could be due to a triangle singularity were done in Ref.~\cite{Guo:2015umn,Liu:2015fea}, but it was shown in Ref.~\cite{guo} that if the quantum numbers of this state are $3/2^-, 5/2^+$, the $\chi_{c1} p$ that merges to form the final $J/\psi p$ state is at threshold and would be in $p$ or $d$ wave respectively. Since the TS appears from placing all intermediate states on shell, the signal coming from the suggested mechanism is drastically reduced and the shape is also distorted such that it cannot reproduce the observed signal. 

   The TS has been discussed in the analysis of some reactions where its consideration can lead to different conclusions than using standard partial wave analysis tools \cite{adam1,adam2,pilloni}.  

     Further examples of TS have recently been investigated. Some of them show that resonances accepted in the PDG \cite{pdg} actually correspond to triangle singularities, which produce a peak, although not related to the interaction of quarks or hadrons, but to the structure of the triangle diagram, tied to the masses of the intermediate states. Apart from the case of the $a_1(1420)$ discussed above, the $f_1(1420)$ peak was shown to  correspond to the $f_1(1285)$ decay into $\pi a_0(980)$, through a TS, and $K^* \bar K$ \cite{Debastiani:2016xgg}. The $f_2(1810)$ peak was also shown to come from a TS involving $K^* \bar K^*$ production, followed by $K^* \to \pi K$ and $\bar K^* K \to a_1(1260)$ \cite{Xie:2016lvs}. Some other times the TS helps building up a particular decay channel of a resonance generated from the interaction of hadrons. This is the case of the $N(1700)$ which is generated from the $\rho N$ interaction with other coupled channels \cite{angelsvec,garzon},
but which gets a sizeable $\pi \Delta$ decay channel through the
mechanism, $N(1700) \to \rho N$ followed by $\rho \to \pi \pi $ and then
$\pi N \to \Delta$ \cite{Roca:2017bvy}. It is also the case of the
$N(1875) (3/2^-)$, which emerges from the interaction of $\Delta \pi$
and $\Sigma^* K$ channels \cite{Sarkar:2004jh}, but that builds up the
$N(1535) \pi$ and $N \sigma$ decay channels from two TS \cite{daris}. 

   In some other cases a TS has been shown to solve some known puzzle, like the enhancement in the $\gamma p \to K^+ \Lambda(1405)$ cross section around $\sqrt s= 2110$~MeV \cite{Moriya:2013hwg} which was discussed from the TS perspective in Ref.~\cite{Wang:2016dtb}, and the $\pi N(1535)$ contribution to the $\gamma p \to \pi^0 \eta p$ reaction \cite{Gutz:2014wit} which was discussed from that perspective in Ref.~\cite{Debastiani:2017dlz}.

    Finally, based on known hadron dynamics, it has become relatively easy to make predictions of peaks that should show up in some reactions, which are solely tied to TS. In this line we can quote the $B^-\rightarrow K^-\pi^-D_{s0}^+$ and $B^-\rightarrow K^-\pi^-D_{s1}^+$ reactions \cite{Sakai:2017hpg}, where 
one finds this type of non-resonant peaks at 2850 MeV in the
invariant mass of $\pi^-D_{s0}^+$ pair and at  3000 MeV in the
invariant mass of $\pi^-D_{s1}^+$  pair respectively  \cite{Sakai:2017hpg}, or the 
$B_c\to B_s \pi\pi$ reaction, which develops a peak 
 at 5777 MeV in the invariant mass of $B_s^0  \pi^+$ \cite{Liu:2017vsf}.

      Coming back to the $f_0(980)-a_0(980)$ mixing, the works done on the subject have shown that the different $K^+$ and $K^0$ masses are responsible for this mixing. In this sense, mechanisms that proceed via a triangle singularity with $K \bar K$ in the intermediate states of the triangle diagram should stress this mixing and make the isospin violating process more efficient. This is because the triangle singularity emerges from having the particles on shell, and this is where the differences of masses play a more relevant role. In this sense, the   $\eta(1405) \to \pi^0 f_0(980)$ reaction is a good example. However, the reaction occurs at a fixed energy, $1405$ MeV. The purpose of the present work is to suggest a reaction where we can change the initial energy to show the isospin violation as a function of the energy, and see that it peaks at the energy where the TS develops. We have found such a case in the $D_s^+ \to \pi^+ \pi^0 a_0(980) (f_0(980))$ reactions which we discuss here. The $D_s^+$ state decays to $\pi^+ s \bar s$ and the $s \bar s$ quarks hadronize to $\bar K^* K$, the $\bar K^*$ decays to $\bar K \pi^0$ and the
$K \bar K$ merge to produce the  $a_0(980)$ or  $f_0(980)$. Since the $s \bar s$ system is in $I=0$ the $\pi^0 a_0(980)$ mode  will be the isospin allowed channel, while the 
$\pi^0 f_0(980)$ mode is the isospin forbidden one. We shall see that both decay modes are enhanced around a $\pi^0 a_0(980)$ or $\pi^0 f_0(980)$ invariant mass of 1420 MeV, but the isospin forbidden channel is more enhanced than the isospin allowed one. Also we can evaluate absolute rates and show that they are well within present measurable range. Since the evaluations are based on the notion that the $f_0(980)$ and $a_0(980)$ resonances are generated from the interaction of coupled channels of pseudoscalar mesons, the rates obtained are tied to this picture and an eventual agreement of the future experiment with the predictions done here would further support this picture for which there is already much phenomenological support \cite{review,sigma,ulfrep}. 

\section{Formalism}
\subsection{The $D_s^+\rightarrow\pi^+K^0\ksbz$ reaction}
\begin{figure}[t]
 \centering
 \includegraphics[width=8.5cm]{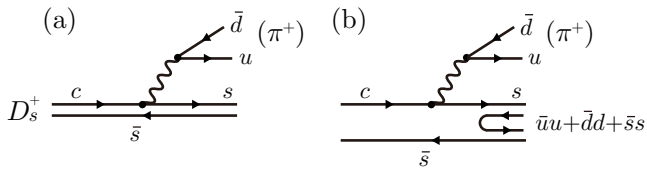}
 \caption{a) Diagrammatic representation of
 $D_s\rightarrow\pi^+\bar{s}s$.\\
 b) Hadronization process through $\bar{q}q$ creation with vacuum quantum
 number.}
 \label{fig1}
\end{figure}

If we look at the $D_s^+$ Cabibbo favored, and color favored decay process
at the quark level, we have the diagram given in Fig.~\ref{fig1} (a),
corresponding to external emission in the classification of
Refs.~\cite{chau1,chau2}.
The process is isospin selective because $s\bar{s}$ has $I=0$.
The $s\bar{s}$ can hadronize with strong interaction leading to two
mesons in $I=0$ incorporating a $\bar{q}q$ pair with the quantum numbers
of the vacuum.
In order to see the meson content of
$s\bar{s}(u\bar{u}+d\bar{d}+s\bar{s})$, we use the arguments of
Refs.~\cite{liang,weakrep} with the $q\bar{q}$ matrix in terms of mesons and we find
\begin{align}
 s\bar{s}(\bar{u}u+\bar{d}d+\bar{s}s)=K^+K^-+K^0\bar{K}^0+...,\label{eq11}
\end{align}
where the points $...$ indicate terms in $\eta$, $\eta'$ which play no
role in the reaction that we study.
The decomposition in Eq.~(\ref{eq11}) has to do about flavor alone, and
what tells us is that we get the $K\bar{K}$ combination in $I=0$
($(K^+,K^0)$ and $(\bar{K}^0,-K^-)$ are the isospin doublets in our
notation).
However, we can get equally $K\ksb$ and this is the channel that we
will pick up to study our process.
Hence we shall look at the decay
\begin{align}
 D_s^+\rightarrow\pi^+(K^+\ksm+K^0\ksbz).\label{eq21}
\end{align}
The reason to choose this channel is that we have the rate for
$D_s^+\rightarrow\pi^+K^{*-}K^+\rightarrow\pi^+\pi^0K^-K^+$ decay
\cite{cleo}.
Since $K^{*-}$ has a branching fraction twice as big for
$\pi^-\bar{K}^0$ than for $\pi^0K^-$, the rate for
$D_s^+\rightarrow\pi^+K^{*-}K^+$ is three times bigger than the one for
$\pi^-\pi^0K^-K^+$ and thus
\begin{align}
 {\rm BR}(D_s^+\rightarrow\pi^+K^{*-}K^+)=3\times(6.37\pm 0.21\pm 0.56)\cdot10^{-2},\label{eq22}
\end{align}
quite a large rate.

In the reaction of Eq.~(\ref{eq21}) angular momentum is conserved and
since we have a vector meson $(J^P=1^-)$ and pseudoscalar meson $(0^-)$
in the final state
%and the rest are pseudoscalar mesons,
we need a $p$-wave.
It is easy to see that non relativistically the right coupling is
$\vec{\epsilon}\,^*_{\ksb}\cdot\vec{p}_{\pi^+}$.
Indeed, the $W^+\pi^+$ vertex goes as $(\partial_\mu\pi^+)W^{+\mu}$
\cite{gasser,stefan} and the $csW$ as $\gamma^\nu(1-\gamma_5)W_\nu$
\cite{chau1,robert}.
The $\gamma^i\gamma_5$ matrix is proportional to $\sigma^i$ at the
quark level which
will be needed to pass from a pseudoscalar to a vector and we are left
with the $\partial_i\pi^+$ component.
Hence, we take
\begin{align}
 t_{D_s^+\rightarrow\pi^+K^+K^{*-}}=C\,\polksm\cdot\vec{p}_{\pi^+},\label{eq31}
\end{align}
and we shall take $C$ constant since there is not much phase space for
this reaction.
When evaluating the triangle diagram we shall work in the $\bar{K}^*K$
system at rest where the $\bar{K}^*$ has a small three momentum.
This is also the case in the $\pi^+K^0\bar{K}^{*0}$ reaction and we
neglect the $\epsilon^0(\bar{K}^*)$ component in Eq.~(\ref{eq31}), but
evaluate $\vec{p}_{\pi^+}$ in the $K\bar{K}^*$ rest frame.

Since we shall need the constant $C$ in the evaluation of the triangle
diagram we proceed to its evaluation by using Eqs.~(\ref{eq22}) and
(\ref{eq31}).
Summing over the polarization of the $K^{*-}$,
we have for the $D_s^+\rightarrow\pi^+K^{*-}K^+$ reaction,
\begin{align}
 \frac{d\Gamma_{\dspkks}}{d\minv(K^+K^{*-})}=&\frac{1}{(2\pi)^3}\frac{p_{\pi^+}\tilde{p}_{K^{*-}}}{4m_{D_s^+}^2}C^2p'^2_{\pi^+},\label{eq32}
\end{align}
where $p_{\pi^+}$ is the $\pi^+$ momentum in the $D_s^+$ rest frame,
$\tilde{p}_{K^{*-}}$ the one of the $K^{*-}$ in the $K^+K^{*-}$ rest
frame and $p'_{\pi^+}$ the $\pi^+$ momentum in the latter frame.
These momenta are given by
\begin{align}
 p_{\pi^+}=&\frac{\lambda^{1/2}(m_{D_s^+}^2,m_{\pi^+}^2,\minv^2(K^+K^{*-}))}{2m_{D_s^+}},\\
 \tilde{p}_{K^{*-}}=&\frac{\lambda^{1/2}(\minv^2(K^+K^{*-}),m_{K^+}^2,m_{\ksm}^2)}{2\minv(K^+K^{*-})}, \\
 p'_{\pi^+}=&\frac{\lambda^{1/2}(m_{D_s^+}^2,m_{\pi^+}^2,\minv^2(K^+K^{*-}))}{2\minv(K^+K^{*-})}, 
\end{align}
where $\lambda(x,y,z)$ is the K\"allen function defined by
$\lambda(x,y,z)=x^2+y^2+z^2-2xy-2yz-2zx$. 

\subsection{The $\dsppa$ $(f_0(980))$ reactions}
\begin{figure}[t]
 \centering
 \includegraphics[width=8.5cm]{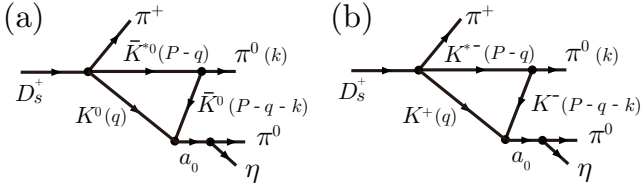}
 \caption{Triangle mechanism which produces $\pi^+\pi^0a_0(980)$.
 The $\pi^+\pi^0f_0(980)$ channel could be seen replacing $\pi^0\eta$ by
 $\pi^+\pi^-$ at the end.
 The momenta of the particles are given in the brackets.}
 \label{fig2}
\end{figure}

In order to produce the $a_0(980)$ or $f_0(980)$, we will look at the
decay products $\pi^0\eta$ and $\pi^+\pi^-$ of the $a_0(980)$ and
$f_0(980)$ respectively.
The mechanism to produce the $a_0(980)$ is depicted in Fig.~\ref{fig2}.
The mechanism of Fig.~\ref{fig2} involves a triangle diagram.
The $\bar{K}^*$ decays to $\pi^0\bar{K}$ and then the remaining $K$ and
this $\bar{K}$ fuse to give the $a_0(980)$ or $f_0(980)$.
The sum of the two diagrams is constructive for $\pi^0\pi^0\eta$
production via $\pi^0a_0$ and destructive for $\pi^0\pi^+\pi^-$
production via $\pi^0f_0$.
In the case the $K^+$ and $K^0$ masses are equal, we would have the
$s$-wave $K^+K^-\rightarrow\pi^0\eta$ and $K^0\bar{K}^0\rightarrow
\pi^0\eta$ amplitudes opposite,
but the $K^+K^-\rightarrow\pi^0\eta$ and
$K^0\bar{K}^0\rightarrow\pi^+\pi^-$ equal.
Taking account of the fact that the vertex
$\bar{K}^{*0}\rightarrow\pi^0\bar{K}^0$ has the opposite sign to
$K^{*-}\rightarrow\pi^0K^-$,
%such that
the sum of diagrams in Fig.~\ref{fig2} for $\pi^0f_0$
production (assuming also equal $\bar{K}^{*0}$ and $K^{*-}$ masses
equal) would cancel and we would have exact $I=0$ $(\pi^0a_0(980))$
production, corresponding to the original $\bar{s}s$ state, and no
$I=1$  $(\pi^0f_0(980))$ production.
When the $K$ masses are allowed to have their physical values we get two
sources of isospin symmetry breaking, from the
$K\bar{K}\rightarrow\pi^0\eta$ $(\pi^+\pi^-)$ amplitudes, when they are
evaluated with the actual $K$ masses, and from the loop function of
Fig.~\ref{fig2},
which is different for the two diagrams thanks to the different $K$
masses (also $\bar{K}^*$).
The interesting thing is that we can now tune
the invariant mass of $\pi^0a_0$ $(\pi^0f_0)$ by changing the energy of
the emitted $\pi^+$, and for a certain value of this invariant mass,
we get a triangle singularity that enhances the production of both
$\pi^0a_0$ and $\pi^0f_0$ modes.
The TS will place the $K\ksb\bar{K}$ on shell in the loop integration
when the momenta of the $\ksb$ and $\pi^0$ from the $\ksb$ decay have the
same direction.
Since the different masses of the charged and neutral kaon cause the
$\pi^0f_0(980)$ production, the on-shell contribution is the most
sensitive to these
differences and we expect that the TS will enhance the $\pi^0f_0$
production versus the $\pi^0a_0$ one.

We proceed now to the evaluation of the diagram of Fig.~\ref{fig2}.
Apart from the vertex of Eq.~(\ref{eq31}), we need the
$\ksb\rightarrow\pi\kb$ vertex that are obtained from the ordinary
Lagrangian,
\begin{align}
 \mathcal{L}_{VPP}=&-ig\left<\left[P,\partial_\mu P\right]V^\mu\right>;\hspace{2.5mm}\ g=\frac{m_V}{2f_{\pi}},
\end{align}
where $P$ and $V$ are the ordinary pseudoscalar and vector meson SU(3)
matrices \cite{angelsvec}, $m_V$ the vector mass ($m_V\sim800$ MeV) and
$f_\pi$ the pion decay constant $f_\pi=93$ MeV.
This produces a vertex
\begin{align}
 t_{\ksbz\rightarrow\pi^0\kb^0}=\frac{g}{\sqrt{2}}(p_{\kb^0}-p_{\pi^0})^\mu\epsilon_{\ksbz\mu},
\end{align}
and opposite sign for $\ksm\rightarrow\pi^0K^-$.

With the former ingredients, the amplitude for the diagram of
Fig.~\ref{fig2} (a) is given by
\begin{align}
 t=&\frac{1}{\sqrt{2}}gCt_{K^0\bar{K}^0,\pi^0\eta}\sumpol i\int\frac{d^4q}{(2\pi)^4}\frac{1}{q^2-m_{K^+}^2+i\epsilon}\notag\\
 &\frac{1}{(P-q)^2-m_{\ksbz}^2+i\epsilon}\frac{1}{(P-q-k)^2-m_{K^-}^2+i\epsilon}\notag\\
 &[\polkszb\cdot(2\vec{k}+\vec{q})][\polkszb\cdot\vec{p}_{\pi^+}].\label{eq71}
\end{align}
Summing upon the polarizations of the intermediate vector meson
and taking $\vec{P}=0$, Eq.~(\ref{eq71}) reads
%gives rise to
\begin{align}
 t=&\frac{1}{\sqrt{2}}gCt_{K^0\bar{K}^0,\pi^0\eta}i\int\frac{d^4q}{(2\pi)^4}\frac{1}{q^2-m_{K^+}^2+i\epsilon}\notag\\
 &\frac{1}{(P-q)^2-m_{\ksbz}^2+i\epsilon}\frac{1}{(P-q-k)^2-m_{K^-}^2+i\epsilon}\notag\\
 &\vec{p}_{\pi^+}\cdot(2\vec{k}+\vec{q}),\label{eq72}
\end{align}
Since in the integral of Eq.~(\ref{eq72}) the only vector not integrated
is $\vec{k}$, we use $\int d^3qf(\vec{q},\vec{k})q_j=k_j\int
d^3qf(\vec{q},\vec{k})(\vec{q}\cdot\vec{k})/\vec{k}^2$ with
$f(\vec{q},\vec{k})$ the remaining terms terms in Eq.~(\ref{eq72}), and then we can write
Eq.~(\ref{eq72}) as
\begin{align}
 t=\frac{1}{\sqrt{2}}\,g\,C\,t_{K\kb^0,\pi^0\eta}\,t_T(\vec{p}_{\pi^+}\cdot\vec{k}),\label{eq73}
\end{align}
where $t_T$ is given by
\begin{align}
 t_T=&i\int\frac{d^4q}{(2\pi)^4}\frac{1}{q^2-m_{K^0}^2+i\epsilon}\frac{1}{(P-q)^2-m_{\ksbz}^2+i\epsilon}\notag\\
 &\cdot\frac{1}{(P-q-k)^2-m_{\kb^0}^2+i\epsilon}\left(2+\frac{\vec{q}\cdot\vec{k}}{\vec{k}^2}\right)\label{eq81}
\end{align}
%from where,
Performing analytically the $q^0$ integration in Eq.~(\ref{eq81}), we obtain \cite{guo,acetijorgi}
\begin{widetext}
 \begin{align}
  t_T=&\int\frac{d^3q}{(2\pi)^3}\frac{1}{8\omkz\omkszb\omkzb}\frac{1}{k^0-\omkzb-\omkszb+i\Gamma_{\ksbz}/2}
  \frac{1}{\minv(\pi^0a_0)+\omkz+\omkzb-k^0}\notag\\
  &\cdot\frac{2\minv(\pi^0a_0)\omkz+2k^0\omkzb-2(\omkz+\omkzb)(\omkz+\omkszb+\omkzb)}{[\minv(\pi^0a_0)-\omkz-\omkzb-k^0+i\epsilon][\minv(\pi^0a_0)-\omkszb-\omkz+i\Gamma_{\ksbz}/2]}\left(2+\frac{\vec{q}\cdot\vec{k}}{\vec{k}^2}\right),\label{eq82}
 \end{align}
\end{widetext}
where $\omega_{K^0}=\sqrt{\vec{q}\,^2+m_{K^0}^2}$,
$\omega_{\kb^0}=\sqrt{(\vec{q}+\vec{k})^2+m_{\kb^0}^2}$,
$\omega_{\ksbz}=\sqrt{\vec{q}\,^2+m_{\ksbz}^2}$, $k^0=\frac{\minv^2(\pi^0a_0)+m_{\pi^0}^2-\minv^2(\pi^0\eta)}{2\minv(\pi^0a_0)}$, and $k=\frac{1}{2\minv(\pi^0a_0)}\lambda^{1/2}(\minv^2(\pi^0a_0),m_{\pi^0}^2,\minv^2(\pi^0\eta))$.
In Eq.~(\ref{eq73}), there is information on $\sqrt{s}=P^0$,
$\minv(\pi^0\eta)$ and $\cos\theta$ with $\theta$ the angle between
$\vec{p}_{\pi^+}$ and $\vec{k}$, but in the integral over the phase
space of $|t|^2$, $1/2\int d\cos\theta\cos^2\theta=1/3$, and we can
define a $t_{\rm eff}$ such that
\begin{align}
 |t_{\rm eff}|^2=\frac{1}{3}\vec{p}\,'^2_{\pi^+}\vec{k}\,^2\left|\frac{1}{\sqrt{2}}Cgt_T\,t_{K^0,\bar{K^0},\pi^0\eta}\right|^2,
\end{align}
and then, summing the two diagrams of Fig.~\ref{fig2},
\begin{align}
 \frac{d^2\Gamma}{d\minv(\pi^0a_0)d\minv(\pi^0\eta)}=&\frac{1}{(2\pi)^5}\frac{p_{\pi^+}k\,\tilde{p}_\eta}{4m_{D_s^+}^2}|t'_{\rm
 eff}|^2,\label{eq91}
\end{align}
where $\tilde{p}_\eta$ is the $\eta$ momentum in the $\pi^0\eta$
center-of-mass frame, and
\begin{align}
 |t'_{\rm
 eff}|^2=&\frac{1}{6}C^2g^2p'^2_{\pi^+}k^2\left|t_T(K^0\kb^0\ksbz)t_{K^0\kb^0,\pi^0\eta}\right.\notag\\
 &\left.\hspace{1.5cm}-t_T(K^+K^-\ksm)t_{K^+K^-,\pi^0\eta}\right|^2.\label{eq92}
\end{align}
For the case of $f_0(980)$ production, we use the same Eq.~(\ref{eq92})
substituting $\pi^0\eta$ in $T$ matrices by $\pi^+\pi^-$.
We can see there that since
$t_{K^0\kb^0,\pi^0\eta}=-t_{K^+K^-,\pi^0\eta}$ and
$t_{K^0\kb^0,\pi^+\pi^-}=t_{K^+K^-,\pi^+\pi^-}$ in the strict isospin limit,
%and  in this same limit,
the two terms in Eq.~(\ref{eq92}) add for the case of the $a_0$
production and subtract in the case of the $f_0$ production.
In the strict isospin limit, the two terms cancel for the $f_0$ production,
as it should be.

The integrand of Eq.~(\ref{eq82}) is regularized including the factor
$\theta(q_{\rm max}-|\vec{q}\,^*|)$, where $\vec{q}\,^*$ is the momentum of
the $K$ in the rest frame of $a_0$ $(f_0)$ (see Eq.~(22) of Ref.~\cite{guo}), with
$q_{\rm max}=600$ MeV as it is needed in the chiral unitary approach
that reproduces the $f_0(980)$ and $a_0(980)$ (see Refs.~\cite{liang,dai}).

\section{Results}
\begin{figure}[t]
 \includegraphics[width=8.5cm]{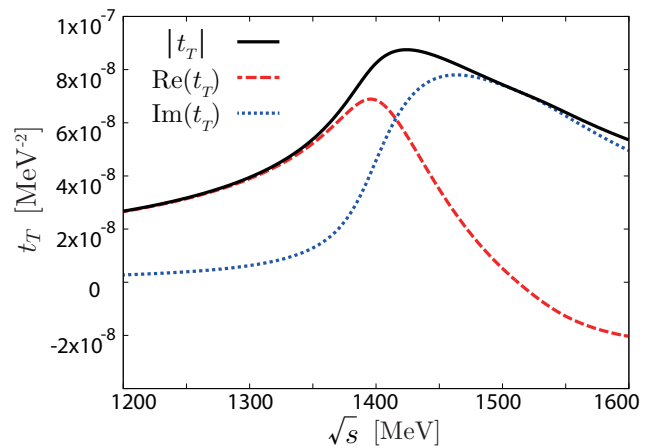}
 \centering
 \caption{Re$(t_T)$, Im$(t_T)$ and $|t_T|$ of Eq.~(\ref{eq82}).}
 \label{fig3}
\end{figure}

In Fig.~\ref{fig3}, we show the results of $t_T$ as a function of
$\sqrt{s}\equiv\minv(\pi^0a_0)$ taking for $\minv(\pi^0\eta)$ (or
$\minv(\pi^+\pi^-)$) the value of 980 MeV.
We can see that the amplitude has a shape similar to a Breit-Wigner with
Re$(t_T)$ and Im$(t_T)$ interchanged ($t_T\sim -it_{BW}$).
Yet the origin of this structure does not come from any particular
interaction, but solely from the analytical structure of the loop function.
We can see that $|t_T|$ has a peak around 1420 MeV and its origin is the
triangle singularity developed by the amplitude.
Indeed, according to Ref.~\cite{guo} the diagrams of Fig.~\ref{fig2}
develops a singularity where in the $d^4q$ integration the $K^0\ksbz$
are placed on shell simultaneously, as well as the $K^0\kb^0$, and the angle
between the $\ksbz$ and the $\pi^0$ coming from its decay is zero.
Analytically this is given by Eq.~(18) of Ref.~\cite{guo} and $q_{\rm
on}=q_{a-}$.
One can see that this occurs at about 1420 MeV (one must choose the mass
of $a_0$ slightly above $m_K+m_{\kb}$ to have the relationship fulfilled).
However, the actual singularity (a sharp peak) becomes a broad bump, as
seen in Fig.~\ref{fig3}, when we consider explicitly the width of the
$\ksb$ in the integral of $t_T$,
$\omega_{\ksb}\rightarrow\omega_{\ksb}-i\Gamma_{\ksb}/2$ in
Eq.~(\ref{eq82}).

\begin{figure}[t]
 \centering
 \includegraphics[width=8.5cm]{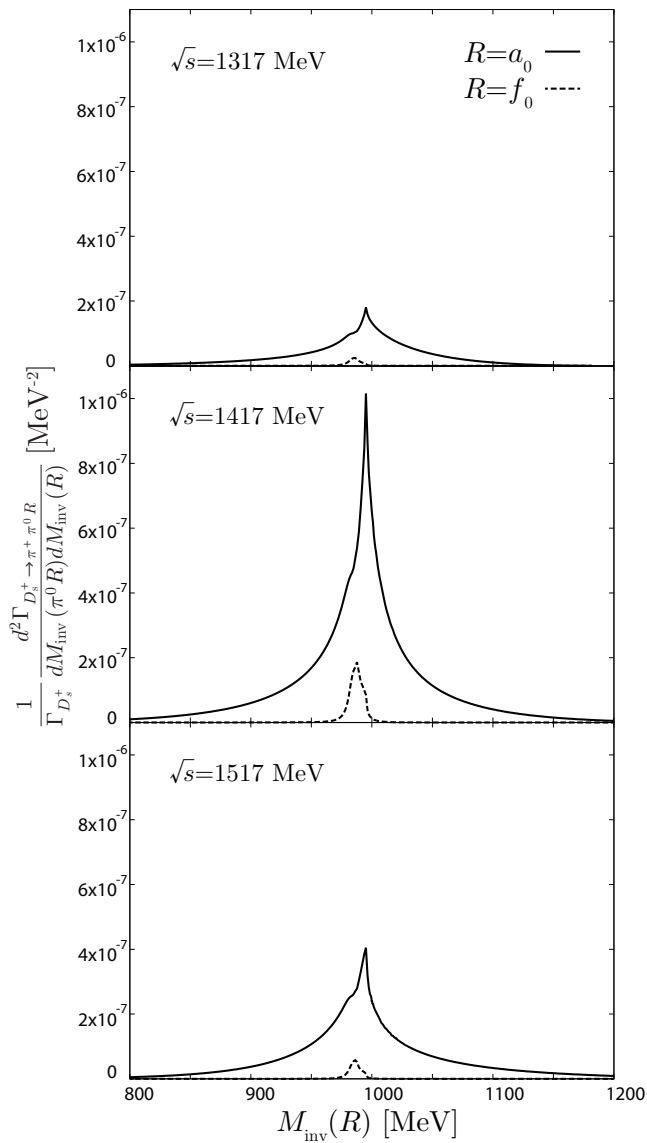}
 \caption{$[d^2\Gamma_{D_s^+\rightarrow\pi^+\pi^0\pi^0\eta}/d\minv(\pi^0a_0)d\minv(\pi^0\eta))]/\Gamma_{D_s^+}$
 and $[d^2\Gamma_{D_s^+\rightarrow\pi^+\pi^0\pi^+\pi^-}/d\minv(\pi^0f_0)d\minv(\pi^+\pi^-))]/\Gamma_{D_s^+}$ as
 functions of $\minv(\pi^0\eta)$ or $\minv(\pi^+\pi^-)$ for fixed value of
 $\minv(\pi^0a_0)$ or $\minv(\pi^0f_0)$ as 1317, 1417, and 1517 MeV, respectively.
 $\minv(R)$ for $R=a_0$ $(f_0)$ means $\minv(\pi^0\eta)$ $(\minv(\pi^+\pi^-))$.}
 \label{fig4}
\end{figure}

In Fig.~\ref{fig4}, we show the results of Eq.~(\ref{eq91}) for
$[d^2\Gamma_{D_s^+\rightarrow\pi^+\pi^0\pi^0\eta}/d\minv(\pi^0a_0)d\minv(\pi^0\eta))]/\Gamma_{D_s^+}$ or 
$[d^2\Gamma_{D_s^+\rightarrow\pi^+\pi^0\pi^+\pi^-}/d\minv(\pi^0f_0)d\minv(\pi^+\pi^-))]/\Gamma_{D_s^+}$ as a
function of $\minv(\pi^0\eta)$ or $\minv(\pi^+\pi^-)$
%for fixed value of $\minv(\pi^0a_0)$ or $\minv(\pi^0f_0)$.
with a fixed value of $\minv(\pi^0a_0)$ or $\minv(\pi^0f_0)$ at 1317,
1417 and 1517 MeV.
For this we have used Eqs.~(\ref{eq22}) and (\ref{eq32}) to determine $C^2$.
%, where the triangle singularity appears.
%We plot the results for these values of $\minv(\pi^0a_0)$ or
%$\minv(\pi^0f_0)$ around the peak of the singularity at
%$\minv(\pi^0a_0)\sim 1420$ MeV.
What we see in the figure is that we get two peaks, corresponding to
the typical $\pi^0\eta$ mass distribution of the $a_0(980)$ and the
$\pi^+\pi^-$ mass distribution of the $f_0(980)$.
The $a_0$ peaks around 995 MeV and the $f_0(980)$ around $985$ MeV.
We also observe a larger strength for $\pi^0a_0$ production (isospin
allowed mode) than for the $\pi^0f_0$ production (isospin suppressed mode).
However, the amount of the $\pi^0f_0$ production is sizable.
The strength of the two distributions at the respective peaks for
$\minvpa$ ($\minvpf$) at $\sqrt{s}=1417$ MeV
is about 16\% for $\pi^0f_0$ versus $\pi^0a_0$, a sizable isospin
violation.
We also see that when we change $\minvpa$ ($\minvpf$) by
100 MeV up and down from this middle energy the strength of both
distributions is sizably decreased.
The maximum strength corresponds to $\minvpa$ $(\minvpf)\sim 1420$ MeV
where the peak of the singularity of the triangle diagram appears.
We also observe that the relative weight of the peaks $\pi^0f_0$ and
$\pi^0a_0$ is decreased by about a factor of two, indicating that the
maximum of the isospin violation appears at the $\minvpa$
$(\minvpf)$ where we have the peak of the triangle singularity.
It should be noted that, although one could interpret this as a
$a_0-f_0$ mixing we have deliberately avoided this perspective and
independently have calculated the rate for $\pi^0f_0$ and $\pi^0a_0$
production.
The isospin violation ($\pi^0f_0$ production) is possible because the
$t_{K\kb,\pi^0\eta}$ and $t_{K\kb,\pi^+\pi^-}$ amplitudes already
contain isospin symmetry breaking terms as soon as the chiral unitary
approach is implemented with different masses of the kaons.
The second reason is the loop of the triangle diagram that also induces
isospin violation from the different masses of the kaons and the $K^*$.
We have checked that the most important source for this isospin breaking
comes from the triangle singularity.

We should also note that we have not explicitly used the $f_0(980)$ and
$a_0(980)$ resonances in the approach.
They are dynamically generated by the $\pi^0\eta$, $\pi\pi$, $K\kb$,
$\eta\eta$ channels \cite{npa,kaiser,markushin,juanito,liang,dai} and they
are implicitly contained in the $t_{K\kb,\pi^0\eta}$ and
$t_{K\kb,\pi^+\pi^-}$ amplitudes.
Note that the apparent width of the $f_0(980)$ distribution is about
10~MeV much narrower than the $f_0$ natural width of about $30-50$~MeV,
because as discussed in Refs.~\cite{crisreport,Aceti:2012dj},
the width of the isospin
violating distribution is of the order of the magnitude of the difference
of the $K^+$ and $K^0$ masses.
This was seen clearly in the experiment in the
$\eta(1405)\rightarrow\pi^0f_0(980)$ \cite{BESIII:2012aa} and one should not take
this width as a measure of the $f_0(980)$ width, which should be looked at
in isospin allowed processes.

As we have seen, the amount of isospin violating $\pi^0f_0$ production is
a function  of $\minv(\pi^0f_0)$ and hence, as already discussed in
Ref.~\cite{Aceti:2012dj}, the concept of a universal $a_0-f_0$ mixing parameter is not
an appropriate one.
It is better to talk in terms of $a_0$ isospin allowed and $f_0$ isospin
forbidden production, or vice versa, which depend on the particular
experiment, and even for the same experiment on the particular part of
the phase space chosen, as we have seen here.

\begin{figure}[t]
 \centering
 \includegraphics[width=8.5cm]{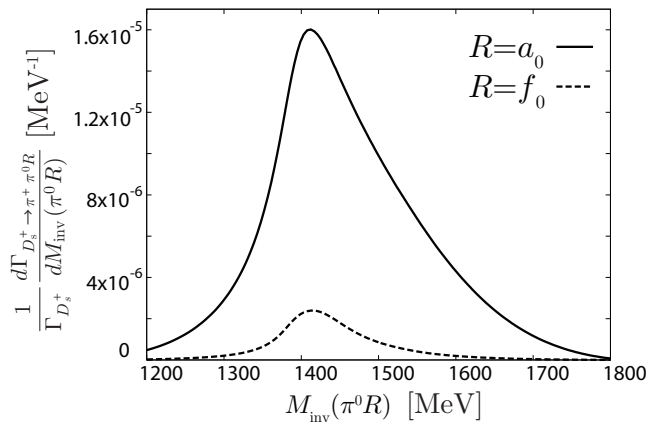}
 \caption{$d\Gamma/d\minv(\pi^0a_0)$ and $d\Gamma/d\minv(\pi^0f_0)$
 integrated over the respective $a_0$ and $f_0$ mass distributions (see
 text).
 Only the $\pi^+\pi^-$ mode of $f_0$ and $\pi^0\eta$ mode of $a_0$ are
 considered here.}
 \label{fig5}
\end{figure}

In order to give a perspective of the amount of isospin violation as a
function of $\minv(\pi^0a_0)$ $(\minv(\pi^0f_0))$, we apply the
following criteria.
The $f_0(980)$ production has a narrow range and we integrate its
strength between $\minv(\pi^+\pi^-)\in [970\mev,1000\mev]$.
The $\pi^0\eta$ mass distribution around the $a_0(980)$ has the typical
cusp form \cite{rubin,midhalo} and has a broad distribution.
Yet it is customary experimentally not to associate the whole strength to the
$a_0(980)$ but subtract a smooth background (note that the amplitudes
of the chiral unitary approach are for $K\kb\rightarrow\pi^0\eta$ and
contain background and pole contributions simultaneously).
In Ref.~\cite{dai} a smooth background was constructed adjusting a
phase space distribution to the sides of the $\pi^0\eta$ distribution,
such that the apparent width of the $a_0$ is about $70-80$ MeV, in the middle
of $50-100$ MeV of the PDG \cite{pdg}.
Then, the strength of the ``$a_0$'' was about one third of the strength
integrated from $\minv(\pi^0\eta)\in [700\mev,1200\mev]$, (see Fig.~3
of Ref.~\cite{dai}).
Then, in Fig.~\ref{fig5} we plot the strength of the integrated mass
distributions of $\pi^0\eta$ and $\pi^+\pi^-$ with this criterion.
We can see in Fig.~\ref{fig5} that both the $\pi^0a_0$ and $\pi^0f_0$
strength peak around $\minv(\pi^0a_0)\sim 1420$ MeV as a consequence of
the TS.

\begin{figure}[t]
 \centering
 \includegraphics[width=8.5cm]{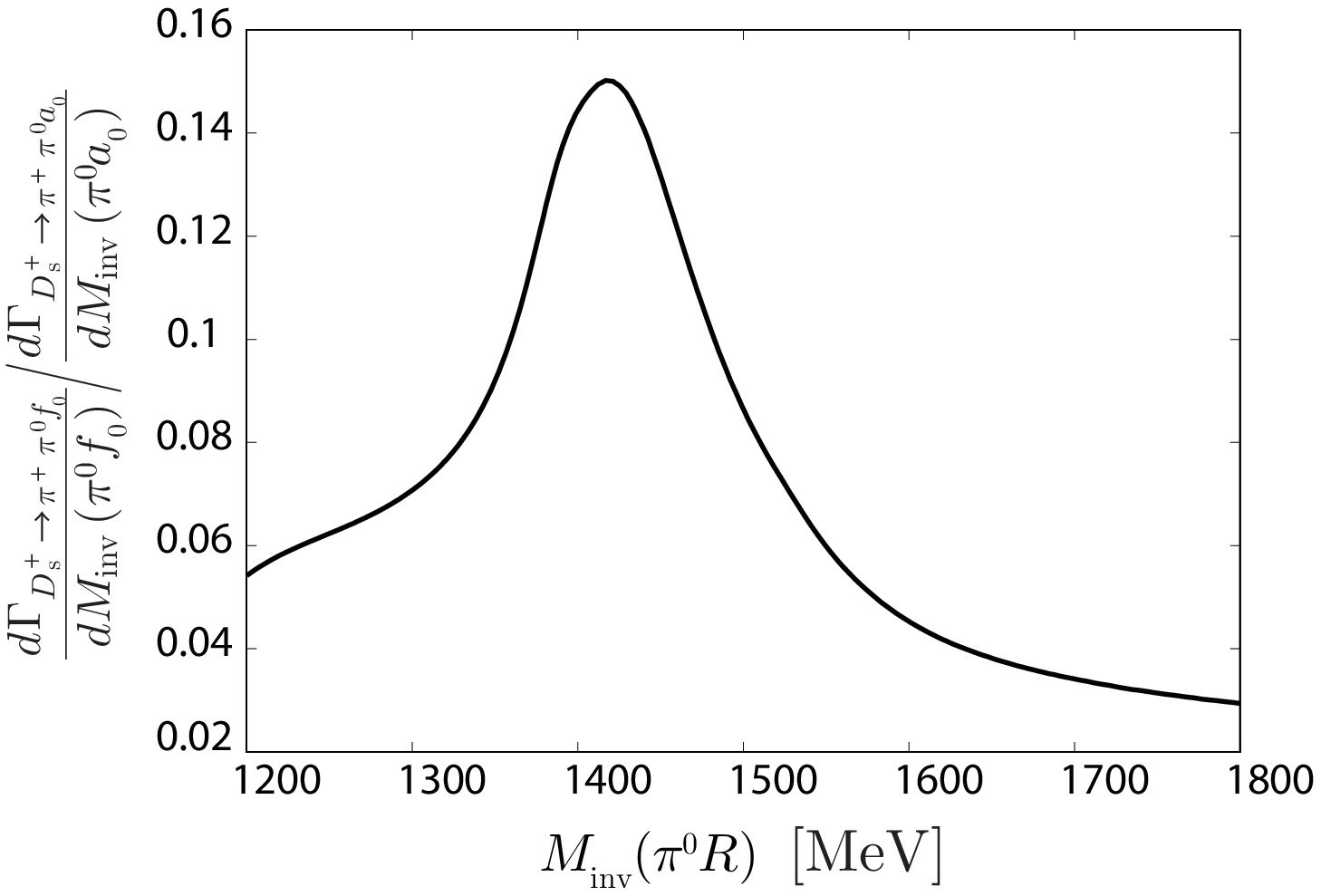}
 \caption{Ratio of $d\Gamma/d\minv(\pi^0a_0)$ and
 $d\Gamma/d\minv(\pi^0f_0)$ as a function of $\minv(\pi^0R)$
 $(R=f_0,a_0)$.}
 \label{fig6}
\end{figure}

In Fig.~\ref{fig6}, we plot the ratio of $d\Gamma/d\minv(\pi^0f_0)$ and
$d\Gamma/d\minv(\pi^0a_0)$.
We see in Fig.~\ref{fig6} that the ratio of $f_0$ to $a_0$ production is
strongly dependent on the $\pi^0R$ $(R=f_0,a_0)$ invariant mass.
By going 100 MeV above and below the peak, the ratio decreases by about
a factor of two and keeps decreasing as we go further away from the
peak.
As we can see, the TS has acted as a magnifier for the isospin violating
$\pi^0f_0$ production process, as we had anticipated.

Finally, we would like to give numbers for the integrated rates of
$\pi^+\pi^0f_0$ and $\pi^+\pi^0a_0$ production by integrating
$d\Gamma/d\minv(\pi^0R)$ in the range of invariant masses of
Fig.~\ref{fig5}.
We find the numbers
\begin{equation}
 \begin{split}
  \br(D_s^+\rightarrow\pi^+\pi^0\, ``f_0\textquotedblright)=&(3.28\pm 0.31)\times10^{-4},\\
  \br(D_s^+\rightarrow\pi^+\pi^0\, ``a_0\textquotedblright)=&(3.28\pm 0.31)\times10^{-3},
 \end{split}\label{eq171}
\end{equation}
which are within present measurable range.
Note that the numbers of Eq.~(\ref{eq171}) are not for the full $f_0$ and
$a_0$ production.
Indeed, in the PDG \cite{pdg} we have
$\Gamma_{K\kb}/\Gamma_{\pi^0\eta}=0.183$.
Also the $f_0$ decays into $\pi^+\pi^-$ and $\pi^0\pi^0$ and
$\Gamma_{\pi^+\pi^-}=2\Gamma_{\pi^0\pi^0}$.
Hence, to correct for that we must divide the ``$a_0$'' production by
$0.85$ and multiply the $f_0$ production by $3/2$.
With this, the numbers of Eq.~(\ref{eq171}) become
\begin{equation}
 \begin{split}
  \br(D_s^+\rightarrow\pi^+\pi^0f_0)=&(4.91\pm 0.46)\times10^{-4},\\
  \br(D_s^+\rightarrow\pi^+\pi^0a_0)=&(3.85\pm 0.36)\times10^{-3}.
 \end{split}\label{eq181}
\end{equation}
The errors in Eqs.~(\ref{eq171}) and (\ref{eq181}) come solely from the
experimental errors in the evaluation of $C$ via Eq.~(\ref{eq22})
summing the errors in quadrature, but they can easily be double of it
accepting similar theoretical errors from different sources, as done in
other examples \cite{liang,dai,etac}.

There is another point worth making.
In Fig.~\ref{fig3}, we see that both Re$(t_T)$ and Im$(t_T)$ have a
peak.
This is a bit different from other cases, where only one of these parts
of $t_T$ have a peak, but not the two
\cite{Roca:2017bvy,daris,Debastiani:2017dlz}.
The present case resembles more the one of Ref.~\cite{Sakai:2017hpg},
where one peak was associated to a threshold and the other
one to a triangle singularity.
In the present case, the peak of Re$(t_T)$ appears because of the
$\bar{K}^{*0}K^0$ threshold, while the one of Im$(t_T)$ comes from the
triangle singularity.
Yet, by looking at Fig.~\ref{fig3} and the mass distribution of
Fig.~\ref{fig5}, it is clear that around the peak of the distributions
most of the strength comes from the triangle singularity.

\section{Conclusions}
The abnormal isospin violation observed in the $\eta(1405) \to \pi^0 f_0(980)$ reaction 
\cite{BESIII:2012aa} and its interpretation as the consequence of a triangle singularity 
in Refs.~\cite{Wu:2011yx,Aceti:2012dj}, prompted us to dig into the problem
looking for a reaction where the energy to produce  $\pi^0 f_0(980)$
could be changed at will. This would allow us to see if, indeed, the TS
has a clear effect enhancing the isospin violation close to the the peak
of the singularity. We found such a reaction in $D_s \to \pi^+ \pi^0
a_0(980) (f_0(980))$, where the freedom to change the energy of the
$\pi^+$ allows one to change the invariant mass of the $\pi^+ \pi^0
a_0(980) (f_0(980))$ system and investigate the amount of isospin
breaking as a function of this invariant mass. The reaction allows to
get a range of $\pi^+ \pi^0 a_0(980) (f_0(980))$ invariant masses that
passes through 1420 MeV, the energy where the triangle mechanism  $D_s^+
\to \bar K^* K$, followed by  $\bar K^* \to \bar K \pi^0$ and the
further merging of
$K \bar K$  to produce the  $a_0(980)$ or  $f_0(980)$ has a triangle
singularity. We could see that, indeed, the isospin violating process of
$D_s \to \pi^+ \pi^0 f_0(980)$ was enhanced versus the isospin allowed
$D_s \to \pi^+ \pi^0 a_0(980)$ as one passed thought the TS peak. This
is due to the fact that the isospin violating reaction was made possible
by the different masses of $K^+$ and $K^0$, and these differences are
stressed by the triangle singularity that places the intermediate
particles (and here the two kaons) on shell, where the difference of the
masses matters most.

It is curious that a weak reaction that violates isospin in the weak
vertex is chosen to investigate isospin violation due to strong
interactions. However, due to Cabibbo selectivity, color enhancement
and the topology of the weak processes, these weak reactions offer
very good filters of isospin in some cases \cite{weakrev}. This was
the case in the reaction chosen. Indeed, the Cabibbo favored, color
favored mode of $D_s$ decay is $D_s^+ \to \pi^+ s \bar s$, and the
$s \bar s$ system has $I=0$. After hadronization with $\bar q q $
pairs, the emerging $K \bar K^*$ state will be in $I=0$. In our
picture, in which the $f_0(980)$ and  $a_0(980)$ resonances are
dynamically generated by the interaction of pseudoscalar mesons, it
is this interaction in the final state which is responsible for the
isospin violation. Since we prove that the isospin mixing depends on
the reaction and for a same reaction like the present one, depends
on the region of the phase space chosen, we deliberately chose not
to talk about $f_0(980)-a_0(980)$ mixing, and the mixing parameter,
because it is not a universal magnitude. It is better to talk in
terms of independent $f_0(980)$ or $a_0(980)$ production and then
investigate the amount of isospin violation. There is mixing of the
two resonances but this in encoded in the   
$K \bar K, \pi^0 \eta$ and $K \bar K, \pi^+ \pi^-$ amplitudes and the
loop functions of the triangle mechanism, and thus is very much
dependent on the reaction and regions of phase space.
From our perspective the results obtained have an extra value. While the
enhancement due to the triangle singularity could be reached in
different ways, the strength obtained and its energy dependence is very
much tied to the nature of the $f_0(980)$ and $a_0(980)$
resonances, which we have assumed as dynamically generated from the
interaction of pseudoscalar mesons. The rates obtained are within
present measurable range and we can only encourage experimental teams to
carry out this reaction, which undoubtedly will bring further light into
the issue of  $f_0(980)-a_0(980)$ mixing and the nature of the low mass
scalar mesons.

\section*{Acknowledgments}

This work is partly supported by the Spanish Ministerio
de Economia y Competitividad and European FEDER funds
under the contract number FIS2011-28853-C02-01, FIS2011-
28853-C02-02, FIS2014-57026-REDT, FIS2014-51948-C2-
1-P, and FIS2014-51948-C2-2-P, and the Generalitat Valenciana
in the program Prometeo II-2014/068 (EO). 
This work is also partly supported by the National Natural Science
Foundation of China under Grants No. 11565007 and No. 11647309.


\begin{thebibliography}{99}

%\cite{Achasov:1979xc}
\bibitem{Achasov:1979xc}
  N.~N.~Achasov, S.~A.~Devyanin and G.~N.~Shestakov,
  %``The S* Delta0 Mixing as the Threshold Phenomenon,''
  Phys.\ Lett.\  {\bf 88B} (1979) 367.
  %%CITATION = doi:10.1016/0370-2693(79)90488-X;%%
  %120 citations counted in INSPIRE as of 29 May 2017

%\cite{Hanhart:2003pg}
\bibitem{crisreport} 
  C.~Hanhart,
  %``Meson production in nucleon-nucleon collisions close to the threshold,''
  Phys.\ Rept.\  {\bf 397}, 155 (2004)
  [hep-ph/0311341].
  %%CITATION = doi:10.1016/j.physrep.2004.03.007;%%
  %203 citations counted in INSPIRE as of 29 May 2017

%\cite{Kudryavtsev:2001ee}
\bibitem{Kudryavtsev:2001ee} 
  A.~E.~Kudryavtsev and V.~E.~Tarasov,
  %``On the possibility of observation of a0 - f0 mixing in the p n ---> da0 reaction,''
  JETP Lett.\  {\bf 72}, 410 (2000)
  [Pisma Zh.\ Eksp.\ Teor.\ Fiz.\  {\bf 72}, 589 (2000)]
  [nucl-th/0102053].
  %%CITATION = doi:10.1134/1.1335118;%%
  %22 citations counted in INSPIRE as of 29 May 2017

%\cite{Kerbikov:2000pu}
\bibitem{Kerbikov:2000pu} 
  B.~Kerbikov and F.~Tabakin,
  %``Mixing of the f(0) and a(0) scalar mesons in threshold photoproduction,''
  Phys.\ Rev.\ C {\bf 62}, 064601 (2000)
  [nucl-th/0006017].
  %%CITATION = doi:10.1103/PhysRevC.62.064601;%%
  %30 citations counted in INSPIRE as of 29 May 2017


%\cite{Achasov:2003se}
\bibitem{Achasov:2003se} 
  N.~N.~Achasov and G.~N.~Shestakov,
  %``To search for a0(0)(980) - f0(980) mixing in polarization phenomena,''
  Phys.\ Rev.\ Lett.\  {\bf 92}, 182001 (2004)
  [hep-ph/0312214].
  %%CITATION = doi:10.1103/PhysRevLett.92.182001;%%
  %24 citations counted in INSPIRE as of 29 May 2017

%\cite{Wu:2007jh}
\bibitem{Wu:2007jh} 
  J.~J.~Wu, Q.~Zhao and B.~S.~Zou,
  %``Possibility of measuring a0(980)-f0(980) mixing from J/psi ---> phi a0(980),''
  Phys.\ Rev.\ D {\bf 75}, 114012 (2007)
  [arXiv:0704.3652 [hep-ph]].
  %%CITATION = doi:10.1103/PhysRevD.75.114012;%%
  %45 citations counted in INSPIRE as of 29 May 2017

%\cite{Hanhart:2007bd}
\bibitem{Hanhart:2007bd} 
  C.~Hanhart, B.~Kubis and J.~R.~Pelaez,
  %``Investigation of a0-f0 mixing,''
  Phys.\ Rev.\ D {\bf 76}, 074028 (2007)
  [arXiv:0707.0262 [hep-ph]].
  %%CITATION = doi:10.1103/PhysRevD.76.074028;%%
  %42 citations counted in INSPIRE as of 29 May 2017
	
%\cite{Oller:1997ti}
\bibitem{npa} 
  J.~A.~Oller and E.~Oset,
  %``Chiral symmetry amplitudes in the S wave isoscalar and isovector channels and the $\sigma$, f$_0$(980), a$_0$(980) scalar mesons,''
  Nucl.\ Phys.\ A {\bf 620}, 438 (1997)
  Erratum: [Nucl.\ Phys.\ A {\bf 652}, 407 (1999)]
  [hep-ph/9702314].
  %%CITATION = doi:10.1016/S0375-9474(99)00427-3, 10.1016/S0375-9474(97)00160-7;%%
  %616 citations counted in INSPIRE as of 29 May 2017

%\cite{Oller:1998hw}
\bibitem{ramonet} 
  J.~A.~Oller, E.~Oset and J.~R.~Pelaez,
  %``Meson meson interaction in a nonperturbative chiral approach,''
  Phys.\ Rev.\ D {\bf 59}, 074001 (1999)
  Erratum: [Phys.\ Rev.\ D {\bf 60}, 099906 (1999)]
  Erratum: [Phys.\ Rev.\ D {\bf 75}, 099903 (2007)]
  [hep-ph/9804209].
  %%CITATION = doi:10.1103/PhysRevD.59.074001, 10.1103/PhysRevD.60.099906, 10.1103/PhysRevD.75.099903;%%
  %582 citations counted in INSPIRE as of 29 May 2017



%\cite{Meissner:2000bc}
\bibitem{Meissner:2000bc} 
  U.~G.~Meissner and J.~A.~Oller,
  %``J / psi ---> phi pi pi (K anti-K) decays, chiral dynamics and OZI violation,''
  Nucl.\ Phys.\ A {\bf 679}, 671 (2001)
  [hep-ph/0005253].
  %%CITATION = doi:10.1016/S0375-9474(00)00367-5;%%
  %109 citations counted in INSPIRE as of 29 May 2017

%\cite{Roca:2012cv}
\bibitem{Roca:2012cv} 
  L.~Roca,
  %``Isospin violation in $J/\Psi to \phi \pi^0 \eta$ decay and the $f_0 - a_0$ mixing,''
  Phys.\ Rev.\ D {\bf 88}, 014045 (2013)
  [arXiv:1210.4742 [hep-ph], arXiv:1210.4742 [hep-ph]].
  %%CITATION = doi:10.1103/PhysRevD.88.014045;%%
  %11 citations counted in INSPIRE as of 29 May 2017

%\cite{Roca:2004uc}
\bibitem{Roca:2004uc} 
  L.~Roca, J.~E.~Palomar, E.~Oset and H.~C.~Chiang,
  %``Unitary chiral dynamics in J/Psi ---> VPP decays and the role of scalar mesons,''
  Nucl.\ Phys.\ A {\bf 744}, 127 (2004)
  [hep-ph/0405228].
  %%CITATION = doi:10.1016/j.nuclphysa.2004.08.004;%%
  %49 citations counted in INSPIRE as of 29 May 2017


%\cite{Ablikim:2010aa}
\bibitem{Ablikim:2010aa} 
  M.~Ablikim {\it et al.} [BESIII Collaboration],
  %``Study of $a_0^0(980) - f_0(980)$ mixing,''
  Phys.\ Rev.\ D {\bf 83}, 032003 (2011)
  [arXiv:1012.5131 [hep-ex]].
  %%CITATION = doi:10.1103/PhysRevD.83.032003;%%
  %36 citations counted in INSPIRE as of 29 May 2017


%\cite{Wu:2008hx}
\bibitem{Wu:2008hx} 
  J.~J.~Wu and B.~S.~Zou,
  %``Study a0(0)(980) - f(0)(980) mixing from a0(0)(980) -- -> f(0)(980) transition,''
  Phys.\ Rev.\ D {\bf 78}, 074017 (2008)
  [arXiv:0808.2683 [hep-ph]].
  %%CITATION = doi:10.1103/PhysRevD.78.074017;%%
  %30 citations counted in INSPIRE as of 29 May 2017

%\cite{BESIII:2012aa}
\bibitem{BESIII:2012aa} 
  M.~Ablikim {\it et al.} [BESIII Collaboration],
  %``First observation of $\eta(1405)$ decays into $f_{0}(980)\pi^0$,''
  Phys.\ Rev.\ Lett.\  {\bf 108}, 182001 (2012)
  [arXiv:1201.2737 [hep-ex]].
  %%CITATION = doi:10.1103/PhysRevLett.108.182001;%%
  %52 citations counted in INSPIRE as of 29 May 2017

%\cite{Wu:2011yx}
\bibitem{Wu:2011yx} 
  J.~J.~Wu, X.~H.~Liu, Q.~Zhao and B.~S.~Zou,
  %``The Puzzle of anomalously large isospin violations in $\eta(1405/1475)\to 3\pi$,''
  Phys.\ Rev.\ Lett.\  {\bf 108}, 081803 (2012)
  [arXiv:1108.3772 [hep-ph]].
  %%CITATION = doi:10.1103/PhysRevLett.108.081803;%%
  %70 citations counted in INSPIRE as of 29 May 2017

%\cite{Aceti:2012dj}
\bibitem{Aceti:2012dj} 
  F.~Aceti, W.~H.~Liang, E.~Oset, J.~J.~Wu and B.~S.~Zou,
  %``Isospin breaking and $f_0(980)$-$a_0(980)$ mixing in the $\eta(1405) \to \pi^{0} f_0(980)$ reaction,''
  Phys.\ Rev.\ D {\bf 86}, 114007 (2012)
  [arXiv:1209.6507 [hep-ph]].
  %%CITATION = doi:10.1103/PhysRevD.86.114007;%%
  %30 citations counted in INSPIRE as of 29 May 2017


%\cite{Wu:2012pg}
\bibitem{Wu:2012pg} 
  X.~G.~Wu, J.~J.~Wu, Q.~Zhao and B.~S.~Zou,
  %``Understanding the property of $\eta(1405/1475)$ in the $J/\psi$ radiative decay,''
  Phys.\ Rev.\ D {\bf 87}, no. 1, 014023 (2013)
  [arXiv:1211.2148 [hep-ph]].
  %%CITATION = doi:10.1103/PhysRevD.87.014023;%%
  %27 citations counted in INSPIRE as of 29 May 2017

%\cite{Landau:1959fi}
\bibitem{landau} 
  L.~D.~Landau,
  %``On analytic properties of vertex parts in quantum field theory,''
  Nucl.\ Phys.\  {\bf 13}, 181 (1959).
  %%CITATION = doi:10.1016/0029-5582(59)90154-3;%%
  %348 citations counted in INSPIRE as of 29 May 2017


%\cite{Coleman:1965xm}
\bibitem{Coleman:1965xm} 
  S.~Coleman and R.~E.~Norton,
  %``Singularities in the physical region,''
  Nuovo Cim.\  {\bf 38}, 438 (1965).
  %%CITATION = doi:10.1007/BF02750472;%%
  %166 citations counted in INSPIRE as of 29 May 2017


%\cite{Bayar:2016ftu}
\bibitem{guo} 
  M.~Bayar, F.~Aceti, F.~K.~Guo and E.~Oset,
  %``A Discussion on Triangle Singularities in the $\Lambda_b \to J/\psi K^{-} p$ Reaction,''
  Phys.\ Rev.\ D {\bf 94}, no. 7, 074039 (2016)
  [arXiv:1609.04133 [hep-ph]].
  %%CITATION = doi:10.1103/PhysRevD.94.074039;%%
  %11 citations counted in INSPIRE as of 29 May 2017



%\cite{Liu:2015taa}
\bibitem{qzhao} 
  X.~H.~Liu, M.~Oka and Q.~Zhao,
  %``Searching for observable effects induced by anomalous triangle singularities,''
  Phys.\ Lett.\ B {\bf 753}, 297 (2016)
  [arXiv:1507.01674 [hep-ph]].
  %%CITATION = doi:10.1016/j.physletb.2015.12.027;%%
  %30 citations counted in INSPIRE as of 29 May 2017

%\cite{Adolph:2015pws}
\bibitem{Adolph:2015pws} 
  C.~Adolph {\it et al.} [COMPASS Collaboration],
  %``Observation of a New Narrow Axial-Vector Meson $a_1$(1420),''
  Phys.\ Rev.\ Lett.\  {\bf 115}, no. 8, 082001 (2015)
  [arXiv:1501.05732 [hep-ex]].
  %%CITATION = doi:10.1103/PhysRevLett.115.082001;%%
  %42 citations counted in INSPIRE as of 29 May 2017

%\cite{Ketzer:2015tqa}
\bibitem{mikha} 
  M.~Mikhasenko, B.~Ketzer and A.~Sarantsev,
  %``Nature of the $a_1(1420)$,''
  Phys.\ Rev.\ D {\bf 91}, no. 9, 094015 (2015)
  [arXiv:1501.07023 [hep-ph]].
  %%CITATION = doi:10.1103/PhysRevD.91.094015;%%
  %33 citations counted in INSPIRE as of 29 May 2017

%\cite{Aceti:2016yeb}
\bibitem{fcadai} 
  F.~Aceti, L.~R.~Dai and E.~Oset,
  %``$a_1(1420)$ peak as the $\pi f_0(980)$ decay mode of the $a_1(1260)$,''
  Phys.\ Rev.\ D {\bf 94}, no. 9, 096015 (2016)
  [arXiv:1606.06893 [hep-ph]].
  %%CITATION = doi:10.1103/PhysRevD.94.096015;%%
  %12 citations counted in INSPIRE as of 29 May 2017

%\cite{Ablikim:2013mio}
\bibitem{Ablikim:2013mio} 
  M.~Ablikim {\it et al.} [BESIII Collaboration],
  %``Observation of a Charged Charmoniumlike Structure in $e^+e^-$ → $π^+π^-$ J/ψ at $\sqrt{s}$ =4.26  GeV,''
  Phys.\ Rev.\ Lett.\  {\bf 110}, 252001 (2013)
  [arXiv:1303.5949 [hep-ex]].
  %%CITATION = doi:10.1103/PhysRevLett.110.252001;%%
  %481 citations counted in INSPIRE as of 29 May 2017


%\cite{Ablikim:2013emm}
\bibitem{Ablikim:2013emm} 
  M.~Ablikim {\it et al.} [BESIII Collaboration],
  %``Observation of a charged charmoniumlike structure in $e^+e^- \to (D^{*} \bar{D}^{*})^{\pm} \pi^\mp$ at $\sqrt{s}=4.26$GeV,''
  Phys.\ Rev.\ Lett.\  {\bf 112}, no. 13, 132001 (2014)
  [arXiv:1308.2760 [hep-ex]].
  %%CITATION = doi:10.1103/PhysRevLett.112.132001;%%
  %216 citations counted in INSPIRE as of 29 May 2017

%\cite{Liu:2013dau}
\bibitem{Liu:2013dau} 
  Z.~Q.~Liu {\it et al.} [Belle Collaboration],
  %``Study of $e^+e^- → π^+ π^- J/ψ$ and Observation of a Charged Charmoniumlike State at Belle,''
  Phys.\ Rev.\ Lett.\  {\bf 110}, 252002 (2013)
  [arXiv:1304.0121 [hep-ex]].
  %%CITATION = doi:10.1103/PhysRevLett.110.252002;%%
  %409 citations counted in INSPIRE as of 29 May 2017

%\cite{Xiao:2013iha}
\bibitem{Xiao:2013iha} 
  T.~Xiao, S.~Dobbs, A.~Tomaradze and K.~K.~Seth,
  %``Observation of the Charged Hadron $Z_c^{\pm}(3900)$ and Evidence for the Neutral $Z_c^0(3900)$ in $e^+e^-\to \pi\pi J/\psi$ at $\sqrt{s}=4170$ MeV,''
  Phys.\ Lett.\ B {\bf 727}, 366 (2013)
  [arXiv:1304.3036 [hep-ex]].
  %%CITATION = doi:10.1016/j.physletb.2013.10.041;%%
  %245 citations counted in INSPIRE as of 29 May 2017

%\cite{Wang:2013cya}
\bibitem{Wang:2013cya} 
  Q.~Wang, C.~Hanhart and Q.~Zhao,
  %``Decoding the riddle of $Y(4260)$ and $Z_c(3900)$,''
  Phys.\ Rev.\ Lett.\  {\bf 111}, no. 13, 132003 (2013)
  [arXiv:1303.6355 [hep-ph]].
  %%CITATION = doi:10.1103/PhysRevLett.111.132003;%%
  %161 citations counted in INSPIRE as of 29 May 2017

%\cite{Liu:2013vfa}
\bibitem{Liu:2013vfa} 
  X.~H.~Liu and G.~Li,
  %``Exploring the threshold behavior and implications on the nature of Y(4260) and Zc(3900),''
  Phys.\ Rev.\ D {\bf 88}, 014013 (2013)
  [arXiv:1306.1384 [hep-ph]].
  %%CITATION = doi:10.1103/PhysRevD.88.014013;%%
  %52 citations counted in INSPIRE as of 29 May 2017


%\cite{Liu:2014spa}
\bibitem{Liu:2014spa} 
  X.~H.~Liu,
  %``Influence of threshold effects induced by charmed meson rescattering,''
  Phys.\ Rev.\ D {\bf 90}, no. 7, 074004 (2014)
  [arXiv:1403.2818 [hep-ph]].
  %%CITATION = doi:10.1103/PhysRevD.90.074004;%%
  %12 citations counted in INSPIRE as of 30 May 2017


%\cite{Aaij:2015tga}
\bibitem{Aaij:2015tga} 
  R.~Aaij {\it et al.} [LHCb Collaboration],
  %``Observation of $J/\psi p$ Resonances Consistent with Pentaquark States in $\Lambda_b^0 \to J/\psi K^- p$ Decays,''
  Phys.\ Rev.\ Lett.\  {\bf 115}, 072001 (2015)
  [arXiv:1507.03414 [hep-ex]].
  %%CITATION = doi:10.1103/PhysRevLett.115.072001;%%
  %392 citations counted in INSPIRE as of 30 May 2017

\bibitem{chinese}
R.~Aaij {\it et al.} [LHCb Collaboration],
  %``Study of the production of $\Lambda_b^0$ and $\overline{B}^0$ hadrons in $pp$ collisions and first measurement of the $\Lambda_b^0\rightarrow J/\psi pK^-$ branching fraction,''
  Chin.\ Phys.\ C {\bf 40} (2016) no.1, 011001.

%\cite{Guo:2015umn}
\bibitem{Guo:2015umn} 
  F.~K.~Guo, U.~G.~Meissner, W.~Wang and Z.~Yang,
  %``How to reveal the exotic nature of the P$_c$(4450),''
  Phys.\ Rev.\ D {\bf 92}, no. 7, 071502 (2015)

%\cite{Liu:2015fea}
\bibitem{Liu:2015fea} 
  X.~H.~Liu, Q.~Wang and Q.~Zhao,
  %``Understanding the newly observed heavy pentaquark candidates,''
  Phys.\ Lett.\ B {\bf 757}, 231 (2016)
  [arXiv:1507.05359 [hep-ph]].
  %%CITATION = doi:10.1016/j.physletb.2016.03.089;%%
  %94 citations counted in INSPIRE as of 30 May 2017

\bibitem{adam1}
 A.~P.~Szczepaniak,
  %``Triangle Singularities and XYZ Quarkonium Peaks,''
  Phys.\ Lett.\ B {\bf 747} (2015) 410.

%\bibitem{adam2}
% A.~P.~Szczepaniak,
%  %``Dalitz plot distributions in presence of triangle singularities,''
%\cite{Szczepaniak:2015hya}
\bibitem{adam2}
  A.~P.~Szczepaniak,
  %``Dalitz plot distributions in presence of triangle singularities,''
  Phys.\ Lett.\ B {\bf 757} (2016) 61
%  doi:10.1016/j.physletb.2016.03.064
  [arXiv:1510.01789 [hep-ph]].
  %%CITATION = doi:10.1016/j.physletb.2016.03.064;%%
  %17 citations counted in INSPIRE as of 12 Jun 2017
	
%\cite{Pilloni:2016obd}
\bibitem{pilloni}
  A.~Pilloni {\it et al.} [JPAC Collaboration],
  %``Amplitude analysis and the nature of the Zc(3900),''
  arXiv:1612.06490 [hep-ph].
  %%CITATION = ARXIV:1612.06490;%%
  %5 citations counted in INSPIRE as of 30 May 2017

\bibitem{pdg}
C. Patrignani et al. (Particle Data Group), Chin. Phys. C, {\bf 40}, 100001 (2016).

%\cite{Debastiani:2016xgg}
\bibitem{Debastiani:2016xgg} 
  V.~R.~Debastiani, F.~Aceti, W.~H.~Liang and E.~Oset,
  %``Revising the $f_1(1420)$ resonance,''
  Phys.\ Rev.\ D {\bf 95}, no. 3, 034015 (2017)
  [arXiv:1611.05383 [hep-ph]].
  %%CITATION = doi:10.1103/PhysRevD.95.034015;%%
  %4 citations counted in INSPIRE as of 30 May 2017

%\cite{Xie:2016lvs}
\bibitem{Xie:2016lvs} 
  J.~J.~Xie, L.~S.~Geng and E.~Oset,
  %``$f_2$(1810) as a triangle singularity,''
  Phys.\ Rev.\ D {\bf 95}, no. 3, 034004 (2017)
  [arXiv:1610.09592 [hep-ph]].
  %%CITATION = doi:10.1103/PhysRevD.95.034004;%%
  %4 citations counted in INSPIRE as of 30 May 2017

%\cite{Oset:2009vf}
\bibitem{angelsvec} 
  E.~Oset and A.~Ramos,
  %``Dynamically generated resonances from the vector octet-baryon octet interaction,''
  Eur.\ Phys.\ J.\ A {\bf 44}, 445 (2010)
  [arXiv:0905.0973 [hep-ph]].
  %%CITATION = doi:10.1140/epja/i2010-10957-3;%%
  %123 citations counted in INSPIRE as of 30 May 2017

%\cite{Garzon:2012np}
\bibitem{garzon} 
  E.~J.~Garzon and E.~Oset,
  %``Effects of pseudoscalar-baryon channels in the dynamically generated vector-baryon resonances,''
  Eur.\ Phys.\ J.\ A {\bf 48}, 5 (2012)
  [arXiv:1201.3756 [hep-ph]].
  %%CITATION = doi:10.1140/epja/i2012-12005-x;%%
  %45 citations counted in INSPIRE as of 30 May 2017

%%\cite{Roca:2017bvy}
%\bibitem{Roca:2017bvy} 
%  L.~Roca and E.~Oset,
%  %``Role of a triangle singularity in the $\pi \Delta$ decay of the $N(1700)(3/2^-)$,''
%  Phys.\ Rev.\ C, in print
%  [arXiv:1702.07220 [hep-ph]].
%  %%CITATION = ARXIV:1702.07220;%%
%  %3 citations counted in INSPIRE as of 30 May 2017

%\cite{Roca:2017bvy}
\bibitem{Roca:2017bvy}
  L.~Roca and E.~Oset,
  %``Role of a triangle singularity in the $\pi \Delta$ decay of the $N(1700)(3/2^-)$,''
  Phys.\ Rev.\ C {\bf 95} (2017) no.6,  065211
%  doi:10.1103/PhysRevC.95.065211
  [arXiv:1702.07220 [hep-ph]].
  %%CITATION = doi:10.1103/PhysRevC.95.065211;%%
  %5 citations counted in INSPIRE as of 06 Jul 2017
	
%\cite{Sarkar:2004jh}
\bibitem{Sarkar:2004jh} 
  S.~Sarkar, E.~Oset and M.~J.~Vicente Vacas,
  %``Baryonic resonances from baryon decuplet-meson octet interaction,''
  Nucl.\ Phys.\ A {\bf 750}, 294 (2005)
  Erratum: [Nucl.\ Phys.\ A {\bf 780}, 90 (2006)]
  [nucl-th/0407025].
  %%CITATION = doi:10.1016/j.nuclphysa.2005.01.006, 10.1016/j.nuclphysa.2006.09.019;%%
  %136 citations counted in INSPIRE as of 30 May 2017

%\cite{Samart:2017scf}
\bibitem{daris}
  D.~Samart, W.~h.~Liang and E.~Oset,
  %``Triangle mechanisms in the build up and decay of the $N^*(1875)$,''
  arXiv:1703.09872 [hep-ph].
  %%CITATION = ARXIV:1703.09872;%%
  %2 citations counted in INSPIRE as of 06 Jul 2017
	
%\cite{Moriya:2013hwg}
\bibitem{Moriya:2013hwg} 
  K.~Moriya {\it et al.} [CLAS Collaboration],
  %``Differential Photoproduction Cross Sections of the $\Sigma^0(1385)$, $\Lambda(1405)$, and $\Lambda(1520)$,''
  Phys.\ Rev.\ C {\bf 88}, 045201 (2013)
  Addendum: [Phys.\ Rev.\ C {\bf 88}, no. 4, 049902 (2013)]
  [arXiv:1305.6776 [nucl-ex]].
  %%CITATION = doi:10.1103/PhysRevC.88.049902, 10.1103/PhysRevC.88.045201;%%
  %52 citations counted in INSPIRE as of 30 May 2017

%\cite{Wang:2016dtb}
\bibitem{Wang:2016dtb} 
  E.~Wang, J.~J.~Xie, W.~H.~Liang, F.~K.~Guo and E.~Oset,
  %``Role of a triangle singularity in the $\gamma p\rightarrow K^+ \Lambda(1405)$ reaction,''
  Phys.\ Rev.\ C {\bf 95}, no. 1, 015205 (2017)
  [arXiv:1610.07117 [hep-ph]].
  %%CITATION = doi:10.1103/PhysRevC.95.015205;%%
  %7 citations counted in INSPIRE as of 30 May 2017

%\cite{Gutz:2014wit}
\bibitem{Gutz:2014wit} 
  E.~Gutz {\it et al.} [CBELSA/TAPS Collaboration],
  %``High statistics study of the reaction $\gamma p\to p\pi^0\eta$,''
  Eur.\ Phys.\ J.\ A {\bf 50}, 74 (2014)
  [arXiv:1402.4125 [nucl-ex]].
  %%CITATION = doi:10.1140/epja/i2014-14074-1;%%
  %40 citations counted in INSPIRE as of 30 May 2017

%\cite{Debastiani:2017dlz}
\bibitem{Debastiani:2017dlz} 
  V.~R.~Debastiani, S.~Sakai and E.~Oset,
  %``Role of a triangle singularity in the $\pi N(1535)$ contribution to $\gamma p \to p \pi^0 \eta$,''
  Phys.\ Rev.\ C, in print
  [arXiv:1703.01254 [hep-ph]].
  %%CITATION = ARXIV:1703.01254;%%
  %1 citations counted in INSPIRE as of 30 May 2017

%\cite{Sakai:2017hpg}
\bibitem{Sakai:2017hpg} 
  S.~Sakai, E.~Oset and A.~Ramos,
  %``Triangle singularities in $B^-\rightarrow K^-\pi^-D_{s0}^+$ and $B^-\rightarrow K^-\pi^-D_{s1}^+$,''
  arXiv:1705.03694 [hep-ph].
  %%CITATION = ARXIV:1705.03694;%%

%\cite{Liu:2017vsf}
\bibitem{Liu:2017vsf} 
  X.~H.~Liu and U.~G.~Meissner,
  %``Generating a resonance-like structure in the reaction $B_c\to B_s \pi\pi$,''
  arXiv:1703.09043 [hep-ph].
  %%CITATION = ARXIV:1703.09043;%%
  %2 citations counted in INSPIRE as of 30 May 2017

%\cite{Oller:2000ma}
\bibitem{review} 
  J.~A.~Oller, E.~Oset and A.~Ramos,
  %``Chiral unitary approach to meson meson and meson - baryon interactions and nuclear applications,''
  Prog.\ Part.\ Nucl.\ Phys.\  {\bf 45}, 157 (2000)
  %%CITATION = doi:10.1016/S0146-6410(00)00104-6;%%
  %230 citations counted in INSPIRE as of 30 May 2017

%\cite{Pelaez:2015qba}
\bibitem{sigma} 
  J.~R.~Pelaez,
  %``From controversy to precision on the sigma meson: a review on the status of the non-ordinary $f_0(500)$ resonance,''
  Phys.\ Rept.\  {\bf 658}, 1 (2016)
  [arXiv:1510.00653 [hep-ph]].
  %%CITATION = doi:10.1016/j.physrep.2016.09.001;%%
  %87 citations counted in INSPIRE as of 30 May 2017

%\cite{Guo:2017jvc}
\bibitem{ulfrep}
  F.~K.~Guo, C.~Hanhart, U.~G.~Meissner, Q.~Wang, Q.~Zhao and B.~S.~Zou,
  %``Hadronic molecules,''
  arXiv:1705.00141 [hep-ph].
  %%CITATION = ARXIV:1705.00141;%%

%\cite{Oset:2016lyh}
\bibitem{weakrev} 
  E.~Oset {\it et al.},
  %``Weak decays of heavy hadrons into dynamically generated resonances,''
  Int.\ J.\ Mod.\ Phys.\ E {\bf 25}, 1630001 (2016)
%  doi:10.1142/S0218301316300010
  [arXiv:1601.03972 [hep-ph]].
  %%CITATION = doi:10.1142/S0218301316300010;%%
  %26 citations counted in INSPIRE as of 01 Jun 2017

%\cite{Chau:1982da}
\bibitem{chau1}
  L.~L.~Chau,
  %``Quark Mixing in Weak Interactions,''
  Phys.\ Rept.\  {\bf 95} (1983) 1.
%  doi:10.1016/0370-1573(83)90043-1
  %%CITATION = doi:10.1016/0370-1573(83)90043-1;%%
  %571 citations counted in INSPIRE as of 01 Jun 2017

%\cite{Chau:1987tk}
\bibitem{chau2}
  L.~L.~Chau and H.~Y.~Cheng,
  %``Analysis of Exclusive Two-Body Decays of Charm Mesons Using the Quark Diagram Scheme,''
  Phys.\ Rev.\ D {\bf 36} (1987) 137.
%  doi:10.1103/PhysRevD.36.137
  %%CITATION = doi:10.1103/PhysRevD.36.137;%%
  %194 citations counted in INSPIRE as of 01 Jun 2017

%\cite{Dedonder:2014xpa}
\bibitem{robert}
  J.-P.~Dedonder, R.~Kaminski, L.~Lesniak and B.~Loiseau,
  %``Dalitz plot studies of $D^0 \to K_S^0 \pi^+ \pi^-$ decays in a factorization approach,''
  Phys.\ Rev.\ D {\bf 89} (2014) no.9,  094018
%  doi:10.1103/PhysRevD.89.094018
  [arXiv:1403.2971 [hep-ph]].
  %%CITATION = doi:10.1103/PhysRevD.89.094018;%%
  %14 citations counted in INSPIRE as of 01 Jun 2017

%\cite{Aceti:2015zva}
\bibitem{acetijorgi}
  F.~Aceti, J.~M.~Dias and E.~Oset,
  %``f$_{1}$(1285) decays into a$_{0}$(980)π$^{0}$, f$_{0}$(980)π$^{0}$ and isospin breaking,''
  Eur.\ Phys.\ J.\ A {\bf 51} (2015) no.4,  48
%  doi:10.1140/epja/i2015-15048-5
  [arXiv:1501.06505 [hep-ph]].
  %%CITATION = doi:10.1140/epja/i2015-15048-5;%%
  %21 citations counted in INSPIRE as of 01 Jun 2017
	
%\cite{Liang:2014tia}
\bibitem{liang}
  W.~H.~Liang and E.~Oset,
  %``$B^0$ and $B^0_s$ decays into $J/\psi$ $f_0(980)$ and $J/\psi$ $f_0(500)$ and the nature of the scalar resonances,''
  Phys.\ Lett.\ B {\bf 737} (2014) 70
%  doi:10.1016/j.physletb.2014.08.030
  [arXiv:1406.7228 [hep-ph]].
  %%CITATION = doi:10.1016/j.physletb.2014.08.030;%%
  %53 citations counted in INSPIRE as of 01 Jun 2017

%\cite{Oset:2016lyh}
\bibitem{weakrep}
  E.~Oset {\it et al.},
  %``Weak decays of heavy hadrons into dynamically generated resonances,''
  Int.\ J.\ Mod.\ Phys.\ E {\bf 25} (2016) 1630001
%  doi:10.1142/S0218301316300010
  [arXiv:1601.03972 [hep-ph]].
  %%CITATION = doi:10.1142/S0218301316300010;%%
  %26 citations counted in INSPIRE as of 01 Jun 2017

%\cite{Onyisi:2013bjt}
\bibitem{cleo}
  P.~U.~E.~Onyisi {\it et al.} [CLEO Collaboration],
  %``Improved Measurement of Absolute Hadronic Branching Fractions of the $D_s^+$ Meson,''
  Phys.\ Rev.\ D {\bf 88} (2013) no.3,  032009
%  doi:10.1103/PhysRevD.88.032009
  [arXiv:1306.5363 [hep-ex]].
  %%CITATION = doi:10.1103/PhysRevD.88.032009;%%
  %26 citations counted in INSPIRE as of 01 Jun 2017

%\cite{Gasser:1984gg}
\bibitem{gasser}
  J.~Gasser and H.~Leutwyler,
  %``Chiral Perturbation Theory: Expansions in the Mass of the Strange Quark,''
  Nucl.\ Phys.\ B {\bf 250} (1985) 465.
%  doi:10.1016/0550-3213(85)90492-4
  %%CITATION = doi:10.1016/0550-3213(85)90492-4;%%
  %3528 citations counted in INSPIRE as of 01 Jun 2017

%\cite{Scherer:2002tk}
\bibitem{stefan}
  S.~Scherer,
  %``Introduction to chiral perturbation theory,''
  Adv.\ Nucl.\ Phys.\  {\bf 27} (2003) 277
  [hep-ph/0210398].
  %%CITATION = HEP-PH/0210398;%%
  %384 citations counted in INSPIRE as of 01 Jun 2017

%\cite{Kaiser:1998fi}
\bibitem{kaiser}
  N.~Kaiser,
  %``pi pi S wave phase shifts and nonperturbative chiral approach,''
  Eur.\ Phys.\ J.\ A {\bf 3} (1998) 307.
%  doi:10.1007/s100500050183
  %%CITATION = doi:10.1007/s100500050183;%%
  %104 citations counted in INSPIRE as of 01 Jun 2017

%\cite{Locher:1997gr}
\bibitem{markushin}
  M.~P.~Locher, V.~E.~Markushin and H.~Q.~Zheng,
  %``Structure of f0 (980) from a coupled channel analysis of S wave pi pi scattering,''
  Eur.\ Phys.\ J.\ C {\bf 4} (1998) 317
%  doi:10.1007/s100529800766, 10.1007/s100520050210
  [hep-ph/9705230].
  %%CITATION = doi:10.1007/s100529800766, 10.1007/s100520050210;%%
  %100 citations counted in INSPIRE as of 01 Jun 2017

%\cite{Nieves:1999bx}
\bibitem{juanito}
  J.~Nieves and E.~Ruiz Arriola,
  %``Bethe-Salpeter approach for unitarized chiral perturbation theory,''
  Nucl.\ Phys.\ A {\bf 679} (2000) 57
%  doi:10.1016/S0375-9474(00)00321-3
  [hep-ph/9907469].
  %%CITATION = doi:10.1016/S0375-9474(00)00321-3;%%
  %158 citations counted in INSPIRE as of 01 Jun 2017

%\cite{Xie:2014tma}
\bibitem{dai}
  J.~J.~Xie, L.~R.~Dai and E.~Oset,
  %``The low lying scalar resonances in the $D^0$ decays into $K^0_s$ and $f_0(500)$, $f_0(980)$, $a_0(980)$,''
  Phys.\ Lett.\ B {\bf 742} (2015) 363
%  doi:10.1016/j.physletb.2015.02.006
  [arXiv:1409.0401 [hep-ph]].
  %%CITATION = doi:10.1016/j.physletb.2015.02.006;%%
  %28 citations counted in INSPIRE as of 01 Jun 2017
	
%\cite{Rubin:2004cq}
\bibitem{rubin}
  P.~Rubin {\it et al.} [CLEO Collaboration],
  %``First observation and Dalitz analysis of the D0 ---> K0(S) eta pi0 decay,''
  Phys.\ Rev.\ Lett.\  {\bf 93} (2004) 111801
%  doi:10.1103/PhysRevLett.93.111801
  [hep-ex/0405011].
  %%CITATION = doi:10.1103/PhysRevLett.93.111801;%%
  %28 citations counted in INSPIRE as of 01 Jun 2017

%\cite{Kornicer:2016axs}
\bibitem{midhalo}
  M.~Ablikim {\it et al.} [BESIII Collaboration],
  %``Amplitude analysis of the $\chi_{c1} \to \eta\pi^+\pi^-$ decays,''
  Phys.\ Rev.\ D {\bf 95} (2017) no.3,  032002
%  doi:10.1103/PhysRevD.95.032002
  [arXiv:1610.02479 [hep-ex]].
  %%CITATION = doi:10.1103/PhysRevD.95.032002;%%
  %3 citations counted in INSPIRE as of 01 Jun 2017

%\cite{Debastiani:2016ayp}
\bibitem{etac}
  V.~R.~Debastiani, W.~H.~Liang, J.~J.~Xie and E.~Oset,
  %``Predictions for $\eta_c \to \eta \pi^+ \pi^-$ producing $f_0(500)$, $f_0(980)$ and $a_0(980)$,''
  Phys.\ Lett.\ B {\bf 766} (2017) 59
%  doi:10.1016/j.physletb.2016.12.054
  [arXiv:1609.09201 [hep-ph]].
  %%CITATION = doi:10.1016/j.physletb.2016.12.054;%%
  %3 citations counted in INSPIRE as of 01 Jun 2017
	
\end{thebibliography}
\end{document}